%% file: main.tex
\documentclass[pra,twocolumn,aps,groupedaddress,longbibliography,superscriptaddress]{revtex4-2}
\usepackage{graphicx}
\usepackage{subfigure}
\usepackage{float}
\usepackage{hyperref}
\hypersetup{
      colorlinks=true,
      citecolor=blue,
      linkcolor=blue,
      urlcolor=blue}
\usepackage{amsmath}
\usepackage{amsfonts}
\usepackage{amssymb}
\usepackage{braket}
\usepackage{bm}
\usepackage{times}

\usepackage{scalerel}
\usepackage{tikz}
\usetikzlibrary{svg.path}

\usepackage{upgreek}
\usepackage{algpseudocode}
\usepackage{algorithm}
\graphicspath{{figures/}}

\newcommand{\brho}{\boldsymbol{\rho}}

\newcommand{\Erbium}{{}^{166}\mathrm{Er}}
\newcommand{\micron}{\mathrm{\upmu m}}
\newcommand{\imicron}{\mathrm{\upmu m^{-1}}}

\newcommand{\Hz}{\mathrm{Hz}}
\newcommand{\seq}{\,{=}\,} 

\newcommand{\ky}{\bar{k}_y}
\newcommand{\edd}{\epsilon_{dd}}
\newcommand{\fpe}{\epsilon_{k_y}}
\newcommand{\seeSup}[1]{, see App.\,#1} 
\DeclareMathOperator\sinc{sinc}
\definecolor{orcidlogocol}{HTML}{A6CE39}
\tikzset{
  orcidlogo/.pic={
    \fill[orcidlogocol] svg{M256,128c0,70.7-57.3,128-128,128C57.3,256,0,198.7,0,128C0,57.3,57.3,0,128,0C198.7,0,256,57.3,256,128z};
    \fill[white] svg{M86.3,186.2H70.9V79.1h15.4v48.4V186.2z}
                 svg{M108.9,79.1h41.6c39.6,0,57,28.3,57,53.6c0,27.5-21.5,53.6-56.8,53.6h-41.8V79.1z M124.3,172.4h24.5c34.9,0,42.9-26.5,42.9-39.7c0-21.5-13.7-39.7-43.7-39.7h-23.7V172.4z}
                 svg{M88.7,56.8c0,5.5-4.5,10.1-10.1,10.1c-5.6,0-10.1-4.6-10.1-10.1c0-5.6,4.5-10.1,10.1-10.1C84.2,46.7,88.7,51.3,88.7,56.8z};
  }
}

\newcommand\orcidicon[1]{\href{https://orcid.org/#1}{\mbox{\scalerel*{
\begin{tikzpicture}[yscale=-1,transform shape]
\pic{orcidlogo};
\end{tikzpicture}
}{|}}}}

\begin{document}
\title{Measurement of the excitation spectrum of a dipolar gas in the macrodroplet regime}
\author{J. J. A. Houwman\,\orcidicon{0009-0003-3342-2322}}
\affiliation{Universit\"at Innsbruck, Institut f\"{u}r Experimentalphysik, 6020 Innsbruck, Austria}
\author{D. Baillie\,\orcidicon{0000-0002-8194-7612}}
\author{P. B. Blakie\,\orcidicon{0000-0003-4772-6514}}
\affiliation{Department of Physics, Centre for Quantum Science, and The Dodd-Walls Centre for \\ Photonic and Quantum Technologies, University of Otago, Dunedin 9016, New Zealand}
\author{G. Natale\,\orcidicon{0000-0002-3726-8196 }}
\author{F. Ferlaino\,\orcidicon{0000-0002-3020-6291}}
\author{M. J. Mark\,\orcidicon{0000-0001-8157-4716}}
\affiliation{Universit\"at Innsbruck, Institut f\"{u}r Experimentalphysik, 6020 Innsbruck, Austria}
\affiliation{Institut f\"{u}r Quantenoptik und Quanteninformation, \"Osterreichische Akademie der \\ Wissenschaften, 6020 Innsbruck, Austria }
\date{\today}
          
\begin{abstract}  	
The excitation spectrum of a cigar-shaped strongly dipolar quantum gas at the crossover from a Bose-Einstein condensate to a trapped macrodroplet is predicted to exhibit peculiar features - a strong upward shift of low momentum excitation energies together with a strong multi-band response for high momenta. By performing Bragg spectroscopy over a wide range of momenta, we observe both key elements and also confirm the predicted stiffening of excitation modes when approaching the macrodroplet regime. Our measurements are in good agreement with numerical calculations taking into account finite size effects.
\end{abstract}
\maketitle
\begin{figure}[!t]
    \centering
    \includegraphics[width=8.6cm]{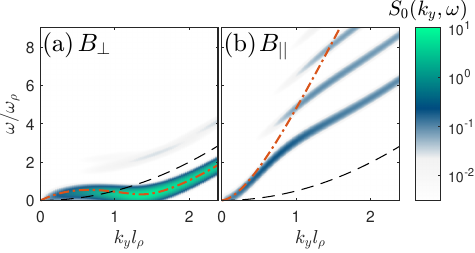}
    \caption{Spectra of excitations of an infinitely elongated dipolar quantum gas in the roton (a) and antiroton regime (b), corresponding to dipoles oriented perpendicular or parallel to the long axis of the system, respectively. The calculations are performed for $\Erbium$ in an infinite tube with transversal trapping frequency $\omega_\rho=2\pi\times198\,\Hz$, a uniform density of $n=2.3\times10^3\imicron$, where $\edd=1.36$ for (a) and $\edd=1.15$ for (b). The color maps shows the strength of the zero-temperature dynamical structure factor $S_0$. The red dashed line is the variational energy (see Eq. \eqref{eq:var_E}) and the black dashed line is the free particle energy $\fpe$. 
    For visibility, a typical experimental Fourier broadening of 47Hz is applied to the DSF.
    \label{fig:infiniteSystem}}
\end{figure}

The successful production of degenerate quantum gases of magnetic lanthanide atoms \cite{Mingwu2011sdb, Aikawa2012bec} has opened the door to study the physics of strongly dipolar Bose-Einstein condensates (BECs), which is fundamentally altered by the long-range and anisotropic dipole-dipole interaction (DDI). The competition between the contact and dipolar interactions, together with a stabilization mechanism based on quantum fluctuations (QF) \cite{Kadau2016otr,FerrierBarbut2016ooq,Waechtler2016qfi,Bisset2016gsp,Baillie2016sbd,Schmitt2016sbd,Chomaz2016qfd}, gives rise to the appearance of a wide range of new exotic states, such as macrodroplets, droplet arrays and supersolids; see Refs.\,\cite{norcia2021dac,chomaz2022dpr} and references therein. \\ \indent
One of the keys to understanding both the stationary and the out-of-equilibrium properties of dipolar phases lies in the knowledge of their spectrum of collective excitations. Remarkably, the spectrum acquires a distinct momentum dependence due to the DDI, which is sensitive to the orientation of the dipoles with respect to the trap axis. As an example, we consider a dipolar BEC confined in a cigar-shaped harmonic trap with the weak axis along $y$. If the dipole orientation is perpendicular to $y$ and the DDI is strong enough, then the spectrum of excitations develops a local minimum at finite momentum \cite{Santos2003rms}, called \textit{roton} in analogy with helium superfluid \cite{Landau1941hto}; see Fig.\,\hyperref[fig:infiniteSystem]{\ref*{fig:infiniteSystem}a}. This phenomenon, recently observed in experiments \cite{Chomaz2018oor,Petter2019ptr}, is a direct consequence of the change of sign of the DDI from mainly repulsive to attractive for increasing momentum $k_y$.
The ratio $\edd = a_{dd}/a_s$, with $a_{dd}$ and $a_s$ being the dipolar and $s$-wave scattering lengths respectively, controls the roton energy gap. By increasing $\edd$, the energy of the roton mode can be decreased until the point where it completely softens, giving rise to the formation of supersolids and independent droplet states\,\cite{chomaz2022dpr}. \\ \indent
In the complementary case, where the dipole moment is aligned along $y$, the DDI is attractive for small $k_y$ and rapidly becomes repulsive as $k_y$ increases. Here, the condensate smoothly enters into a macrodroplet state~\footnote{A macrodroplet state is defined as the ground state of a dipolar gas that exists due to the stabilization of quantum fluctuations, where by contrast mean-field theory (standard GPE) would predict a collapse of the dipolar gas. A macrodroplet can in addition become self-bound by further decreasing the contact interaction.} 
when increasing $\edd$ above unity\,\cite{Chomaz2016qfd,Schmitt2016sbd}.
In this regime, the spectrum of excitations is much less explored. Just recently, theoretical works have revealed remarkable features: The excitation spectrum undergoes a stiffening, i.e. an upward curvature at low $k_y$, and a multi-band response to excitations~\cite{Baillie2017ceo,pal2020enf,pal2022idd}; see Fig.\,\hyperref[fig:infiniteSystem]{\ref*{fig:infiniteSystem}b}. These distinct characteristics, denoted as the \textit{antiroton} effect~\cite{pal2020enf}, call for an experimental verification as they play a crucial role in shaping the properties of macrodroplet states. Specifically, the observed stiffening at low k and the spreading of the system’s response over multiple bands indicates the heightened resilience of a dipolar macrodroplet against excitations. This increased stability could potentially lead to the self-evaporation of these excitations, introducing a novel and intriguing aspect to the behavior of quantum systems bound by beyond-mean-field quantum fluctuations, as seen in dipolar gases and self-bound states in non-dipolar mixtures~\cite{Cabrera2018,Cheiney2018,Semeghini2018,Ferioli2019,Burchianti2020,Boettcher2020}.  \\ \indent
In the present work we report on the experimental observation of the antiroton effect
by measuring the excitation spectrum of an erbium quantum gas across the BEC-macrodroplet crossover. Furthermore, we extend the current theory to finite systems to determine the mechanisms underlying the peculiar shape of the spectrum. 
Not only do we observe the predicted stiffening of the lowest excitation branch at low $k_y$, we also measure a strong multi-band response when increasing the imparted momentum.  \\ \indent
To illustrate the physics of the antiroton, let us first discuss the theoretical framework to determine the excitation spectrum of a dipolar quantum gas in an infinite tube with constant density. The dynamic structure factor (DSF) $S(\mathbf{k},\omega)$ quantifies the density response of a system to a density-coupled scattering probe of momentum $\hbar\mathbf{k}$ and energy $\hbar\omega$ within linear response theory. For a BEC the DSF at $T=0$\:K is given by
\begin{align}\label{eq:DSF}
S_0(\mathbf{k},\omega)=\sum_j\left|\int d\mathbf{x}\,(u_j^*+v_j^*)e^{i\mathbf{k}\cdot\mathbf{x}}\psi_0\right|^2\delta(\hbar\omega-\epsilon_j),
\end{align}
where $\{u_j(\mathbf{x}),v_j(\mathbf{x})\}$ are the Bogoliubov quasi-particle amplitudes with respective energies $\epsilon_j$, and $\psi_0$ is the condensate wavefunction. 
As presented in Refs.\,\cite{pal2020enf,pal2022idd}, the excitations are plane waves along the $y$-direction of the infinite tube i.e.\,$u_{j}(\mathbf{x})\to \mathrm{u}_{\nu,k_y}(\bm{\rho}) e^{ik_y y}$, with $\nu$  labelling the transverse excitation (e.g.\,see \cite{Baillie2017ceo}), $\hbar k_y$ being the $y$-component of momentum and $\bm{\rho}=(x,z)$ the transverse coordinates. 
Together with a variational description of the transverse structure of the condensate and excitations, this Ansatz allows a simple semi-analytical expression
 of the dispersion relation for the lowest excitation branch of the form
\begin{equation} \label{eq:var_E}
\epsilon_{k_y}^\text{var}=\sqrt{\epsilon_{k_y} [ \epsilon_{k_y}+ 2n\tilde{U}_\alpha(k_y) + 3n^{3/2}g_\text{QF}]}.
\end{equation}
Here, $\epsilon_{k_y}=\hbar^2k_y^2/2m$ and the quantities $\tilde{U}_\alpha(k_y)$ and $g_\text{QF}$ describe the (Fourier transform of the) two-body interactions and the effects of quantum fluctuations after the transverse degrees of freedom are integrated out\seeSup{\ref{app:VariationalTheory}}.  \\ \indent
In the roton regime (dipoles perpendicular to $y$), the variational dispersion relation provides an excellent description of the full excitation spectrum, as shown by the remarkable agreement between Eq.\,\eqref{eq:var_E} and the full numerical calculation, see Fig.\,\hyperref[fig:infiniteSystem]{\ref*{fig:infiniteSystem}a}. The latter, described in detail in App.\,\ref{app:Theory} and Ref.\,\cite{Petter2019ptr}, also reveals that the dynamic response of the system essentially involves only the lowest branch.  \\ \indent
In the macrodroplet regime (dipoles parallel to $y$), the situation is very different. As shown in Fig.\,\hyperref[fig:infiniteSystem]{\ref*{fig:infiniteSystem}b}, Eq.\,\eqref{eq:var_E} is only capable to describe the low momentum part ($k_y<1/l_\rho$), where $l_\rho=\sqrt{\hbar/m\omega_\rho}$ is the characteristic transverse length scale and $\omega_\rho$ the transverse trapping frequency. For larger momenta, the variational dispersion relation deviates substantially from the numerics. Moreover, from the full calculation of the DSF, we also observe a strong multi-branch response of the system for $k_y\gtrsim1/l_\rho$ i.e. the DSF for a single momentum features multiple resonant energies with weights on the same order of magnitude. Both aspects suggest the emergence of new phenomena not captured by the above variational method. We find that the same qualitative behavior persists also in the experimentally relevant case of a three-dimensional trapped system. Despite the expected broadening and discretization of the excitation modes, the \textit{antiroton} features are still visible, namely the stiffening of the spectrum with an initial rapid increase in energy with $k_y$ and the strong multi-branch response above $k_y\gtrsim1/l_\rho$.  \\ \indent
\begin{figure}[t]
	\centering
    \includegraphics[width=8.6cm]{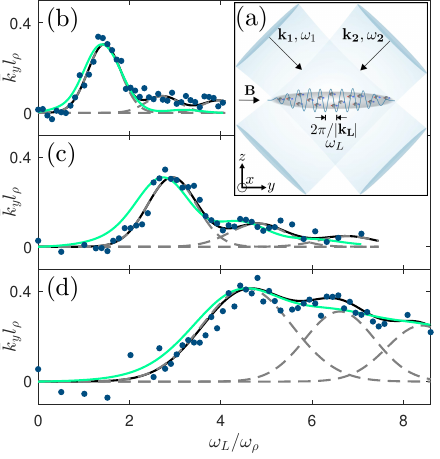}
     \setlength{\belowcaptionskip}{-10pt}
      \setlength{\abovecaptionskip}{-10pt}
     \caption{
     (a) Illustration of the experimental setup: A BEC of $\Erbium$ in an elongated trap with the dipole moment aligned along the weak axis $y$. Bragg spectroscopy is performed by intersecting two beams at the BEC which impart momentum $\hbar\mathbf{k}_L$ and energy $\hbar\omega_L$.
     (b-d)  Measured response of the system $\ky$ as a function of the excitation frequency $\omega_L/\omega_\rho$ for three different values of the imparted momentum $k_L=[0.59,1.0,1.76]/l_\rho$, respectively.
     The gray lines are the sum of three Gaussians (dashed lines) fit to the data. 
     The green lines are the theoretically calculated values of $S_0(k_y,\omega)$, rescaled in amplitude and accounting for the experimental Fourier and Doppler broadening.  
     }
    \label{fig:fig2}
\end{figure} 
\begin{figure*}[t]
	\centering
    \includegraphics[width=\textwidth]{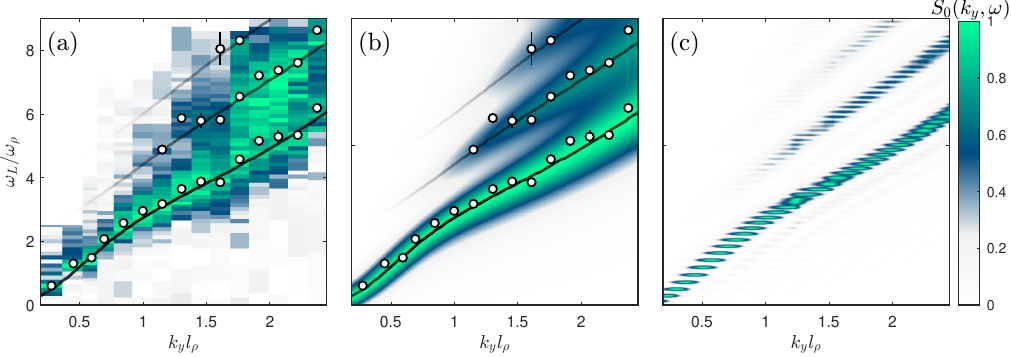}
     \caption{Excitation spectrum of an elongated $\Erbium$ cloud polarised along the weakly confined axis with $\edd=1.13$, $N\seq25(5.0)\times10^4$ and trapping frequencies $\omega_{x,y,z}/2\pi\seq(170,30,230)\mathrm{Hz}$. 
     (a) Measured response of the system $\ky$ as a function of $\omega_L/\omega_\rho$ and $k_L$. The datapoints denote the experimentally detected resonance positions from fits as shown in Fig.\,\hyperref[fig:fig2]{\ref*{fig:fig2}b-d} with the error-bars indicating one standard error from the fit. The black lines represent the maxima of $S_0(k_y,\omega)$ from our numerical calculation.
     (b) Detected resonance positions on top of the calculated $S_0(k_y,\omega)$ as indicated by the colormap, including Fourier and Doppler broadening.
     (c) Calculated $S_0(k_y,\omega)$ for our finite system with minimal broadening. (a-c) Each momentum-column is normalized to the maximum value.
     }
     \label{fig:expSpectrum}
\end{figure*}
To probe the theoretical findings, we experimentally explore the spectrum of excitations of a macrodroplet state using Bragg spectroscopy. We first prepare a dipolar BEC of $\Erbium$ in the lowest Zeeman state at a magnetic field of $\mathbf{B}=1.9\,$G pointing along $z$ in a crossed dipole trap, analogous to Refs.\,\cite{Chomaz2016qfd,Chomaz2018oor}. We then prepare the state of interest by simultaneously rotating the direction of $\mathbf{B}$ to point towards $y$, ramping its magnitude to the desired value, and by reshaping the trap to $\omega_{x,y,z}/2\pi=(170,30,230)\,\mathrm{Hz}$, resulting in a characteristic transverse trapping frequency $\omega_\rho=\sqrt{\omega_x\omega_z}=2\pi\times198\,\Hz$ with a corresponding length scale of $l_\rho=0.55\micron$. Finally, we perform Bragg-spectroscopy \cite{Stenger1999bso} by illuminating the cloud with a moving lattice with wavevector $\mathbf{k}_L=\mathbf{k}_1-\mathbf{k}_2$ and angular frequency $\omega_L = \omega_1 - \omega_2$, resulting in a phase velocity of $v_L=\omega_L/|\mathbf{k}_L|$, see Fig.\,\hyperref[fig:fig2]{\ref*{fig:fig2}a} and App.\,\ref{app:ExpSequence}.  \\ \indent
In the experiment, $\mathbf{k}_L$ is aligned along $y$ such that $\mathbf{k}_L = k_L\hat{\mathbf{y}}$.
The lattice is created by letting two laser beams ($\mathbf{k}_1,\omega_1$) and ($\mathbf{k}_2,\omega_2$), red-detuned by about $40\,\Gamma$ from the main electronic transition of erbium at $401\,$nm, intersect at an angle $\theta$ at the position of the atoms. The two beams drive a two-photon transition when the resonance condition $\hbar \omega_L = \epsilon(k_L)$ is met, transferring momentum $k_L$ and energy $\hbar \omega_L$ to the atoms. The imparted momentum and energy can be independently controlled via the digital micromirror device used to create both beams.
We use a pulse duration of $\tau=8.3\,$ms, corresponding to roughly a quarter of the axial trap period, as a compromise between minimizing Fourier broadening and the influence of the trap on the momentum of the excited atoms.
The power of the Bragg beams is chosen for each measurement in a range of $1-7\,$mW such that a clear excitation signal is seen on resonance while the maximum excited fraction is kept below $20\,$\%. We probe the atomic cloud by standard absorption imaging along $z$ after a total $30\,$ms time-of-flight (TOF) expansion. 
Excitations are then visible either as a separated peak for large $k_L$, or as an asymmetric broadening of the momentum distribution for low $k_L$.
A more detailed description of our Bragg-spectroscopy setup based on a digital micromirror device (DMD) can be found in Ref.\,\cite{Petter2019ptr}. \\ \indent
From the 2D momentum distribution - assuming ballistic expansion during TOF - we extract the momentum profile along $y$ by integrating out the $x$-direction $n(k_y)=\int n(k_x,k_y)dk_x$, and then calculate the mean momentum per particle as $\ky=\int k_yn(k_y)dk_y/\int n(k_y)dk_y$\seeSup{\ref{appendix:imgAnalysis}}. One can show that this quantity is directly proportional to the zero-temperature dynamical structure factor $S_0(\mathbf{k},\omega)$ in the linear response regime as
\begin{equation}
   \ky =  k_L \frac{\pi \tau V_0^2}{2 \hbar}  S_0(k_L,\omega),
\end{equation}
where $V_0$ is the depth of the moving lattice \cite{brunello2001mtt, blakie2002tcb}. This comparison between the mean imparted momentum and the dynamic structure factor is particularly robust since it is insensitive to scattering processes during or after the excitation pulse and does not rely on any fitting parameters\seeSup{\ref{app:CompSandk}}.  \\ \indent
Figure\,\hyperref[fig:fig2]{\ref*{fig:fig2}b-d} shows examples of the measured response of the system $\ky$ as a function of the excitation frequency $\omega_L/\omega_\rho$ for three different values of the imparted momentum $k_L$. At low momentum ($k_L=0.59/l_\rho$), we observe one prominent resonance around $\omega_L\approx1.5\omega_\rho$ indicating the presence of a single excitation branch.
For larger momenta ($k_L=[1.0,1.76]/l_\rho$), we instead observe a multi-peak response, signalizing that additional resonances at higher energies start to appear and become more pronounced when increasing $k_L$, see Fig.\,\hyperref[fig:fig2]{\ref*{fig:fig2}c-d}. A direct comparison of our experimental spectra to the expected response of the system from the full calculation reveals a good agreement with both energies and strengths of the excitation resonances when we take into account the expected Doppler- and Fourier broadening\seeSup{\ref{app:Broadening}}. While the Fourier broadening of $\delta\omega_F/2\pi\approx  47\,\Hz$ does not depend on the probing momentum and determines the resonance width at low $k_L$, the Doppler broadening increases with $k_L$ like $\delta\omega_D/2\pi\approx  58\,\Hz \cdot k_L l_{\rho}$, limiting our ability to observe separated excitation branches due to their increasing spectral width. \\ \indent
We reconstruct the full spectrum of excitations by repeating the above measurements over a wide tuning range of $k_L=[0.06,2.38]/l_\rho$. Figure\,\hyperref[fig:expSpectrum]{\ref*{fig:expSpectrum}} summarizes our experimental results together with the numerical calculations. The measured spectrum of excitations (Fig.\,\hyperref[fig:expSpectrum]{\ref*{fig:expSpectrum}a}) shows the predicted steep upward curvature of the first excitation branch for low $k_L$, with an additional decrease of its slope and the appearance of a strong multi-band response for increasing momentum. The datapoints mark our extracted resonance positions obtained by fitting multiple Gaussians to the obtained spectra for fixed $k_L$ as shown in Fig.\,\hyperref[fig:fig2]{\ref*{fig:fig2}b-d}.  \\ \indent
\begin{figure}[t]
     \centering
     \includegraphics[width=8.6cm]{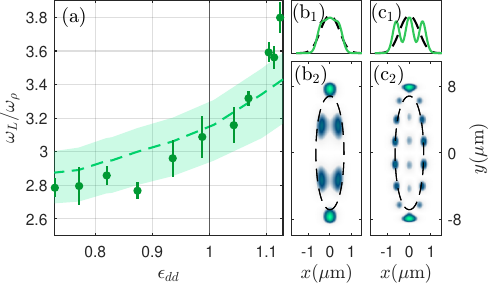}
     \caption{Stiffening and shapes of excitation modes. (a) Frequency of the first excitation branch versus $\edd$ at fixed excitation momentum $k_L = 1.30/ l_\rho$. The error bars denote one standard error from the fit. The shaded area corresponds to the theoretical predictions within a range of atom number $N=[15,35]\times10^4$. (b2,\,c2) Shapes of the numerically calculated excitation modes at $\edd=1.13$ and energies $\epsilon_j/\hbar=0.51\omega_{\rho}$ (b) and $\epsilon_{j'}/\hbar=1.87\omega_{\rho}$ (c) represented via plots of $|\delta n_j|^2 =|(u_j^*+v_j^*)\psi_0|^2$. The dashed lines mark the 1/e size of the condensate. (b1,\,c1) Profiles of the excitation modes after integration along $y$. The dashed lines mark the integrated shape of the condensate.}
    \label{fig:fig4Modes}
\end{figure}
Figure\,\hyperref[fig:expSpectrum]{\ref*{fig:expSpectrum}b} shows our experimental results on top of the calculated DSF taking into account the expected Doppler- and Fourier broadening as direct comparison, while Fig.\,\hyperref[fig:expSpectrum]{\ref*{fig:expSpectrum}c} plots the bare theory spectrum without any broadening as reference. 
We find very good agreement between our experimental results and the numerical calculations, confirming the existence of the antiroton phenomena, which signalizes the macrodroplet's resilience to excitations.
This rigidity is expected to further increase when moving from the BEC towards the macrodroplet regime~\cite{pal2022idd}. This behavior becomes apparent from our study, showing that an ever-increasing energy is needed to excite the system as $\edd$ increases. Figure\,\hyperref[fig:fig4Modes]{\ref*{fig:fig4Modes}a} summarizes these findings. Here, for each point, we experimentally Bragg excite the system at a fixed momentum $k_L = 1.30/l_\rho$ and extract the excitation frequency belonging to the lowest branch of the spectrum.  We then repeat the measurement for various interaction parameters in the range $\edd=[0.73,1.12]$ and observe the expected increase of excitation energy with increasing $\edd$ in quantitative agreement to theory.  \\ \indent
While the results clearly show the stiffening and dispersion of the response of a macrodroplet over multiple branches of excitation, the physical origin of this phenomenon and the reason for the failure of the variational approach still await explanation. We find that a first important indication comes from the theoretical study of the transverse behavior of the system. While the variational approach [Eq.\,\eqref{eq:var_E}] imposes a Gaussian-type transverse profile of the excitation modes, regardless of energy, the numerical results show a marked change in the profile at low and high energy. This observation is exemplified in Fig.\,\hyperref[fig:fig4Modes]{\ref*{fig:fig4Modes}b-c}, in which we compare the profile of the density fluctuations $|\delta n|^2$ for specific excitation modes in the lowest branch at different energies $\epsilon_j$. From the profile integrated along the axial direction, we observe that for increased energy the \textit{transversal} profile of the excitation mode changes from a Gaussian-like shape to a more broadened multi-peak structure, substantially differing from the ground-state wavefunction. This leads to a drastic reduction of the strength of the DSF of the lowest branch, since it scales directly with the overlap integral between the condensate and the excitation wavefunctions, see Eq.\,\eqref{eq:DSF}. As a consequence, the higher branches pick up more relative weight due to the f-sum rule\,\cite{Pitaevskii2016bec}, leading to the multi-band response. This behavior of the shape also explains the breakdown of the variational model, as it assumes a similar transverse structure for the excitations and the condensate which is only valid for small energies in this regime.  \\ \indent
But why do the excitations change their shape in the first place? Here one should be reminded that dipolar interactions exhibit a momentum dependence. In the macrodroplet regime, DDI are becoming attractive for small momenta and strongly repulsive for large momenta. Therefore excitations with large momentum exhibit a large repulsion from the condensate, forcing them to reduce their overlap to lower the energy. This goes hand-in-hand with the transition from collective to single-particle excitations as the excitation energy increases above the chemical potential $\mu$ of the condensate\,\cite{Pitaevskii2016bec}. \\ \indent
In summary, this work presents the first measurement of the excitation spectrum of a dipolar quantum gas in the crossover regime from a Bose-Einstein condensate to a macrodroplet state. While a dominant side-by-side dipole orientation typically leads to a softening of the excitation spectrum, resulting in the emergence of the roton spectrum~\cite{Santos2003rms,Chomaz2018oor} and eventually supersolidity~\cite{chomaz2022dpr}, our observations reveal a contrary behavior. Specifically, we note a stiffening of the excitation spectrum and a spreading of the dynamic structure factor across multiple branches when dipoles are primarily arranged head-to-tail. This phenomenon, referred to as the antiroton effect, becomes particularly pronounced in the macrodroplet phase, imparting a resistance to the formation of density modulations. \\ \indent
Our findings open the path for further investigation into the excitations of the macrodroplet and their impact on the system's behavior, especially in the self-bound regime. In this scenario, it is predicted \cite{Baillie2017ceo,pal2022idd} that reducing the confinement will transform the multiple discrete branches that couple to axial probing into just one discrete branch and a continuum of excitations. The observed increased stiffening is also predicted to induce self-evaporation of excitations~\cite{Petrov2015qms}, prompting questions about interpreting finite temperature effects in such systems. Another unresolved aspect is whether the system has a critical velocity and its relation to the incompressible nature of the droplet~\cite{pal2022idd}. \\ \indent
We acknowledge T.\,Mézières, L.\,Lafforgue, and L.\,Chomaz for discussion and support in the early stages of the experiment. We are grateful to T.\,Bland and E.\,Poli for helpful discussions about the theoretical simulations. The Innsbruck team acknowledges support from the European Research Council through the Advanced Grant DyMETEr (No.\,101054500), the DFG/FWF via FOR 2247/I4317-N36, and NextGeneration EU through AQuSIM (No. FO999896041). A.H. acknowledges funding from the Austrian Science Fund (FWF) within the DK-ALM (No.\,W1259-N27). D.B. and P.B.B. acknowledge support from the Marsden Fund of the Royal Society of New Zealand (MFP-22-UOO-011).

\input{main.bbl}
\input{SuppMat}

\end{document}

%% file: SuppMat.tex
\appendix

\section{Variational model}\label{app:VariationalTheory}

The variational theory for a dipolar BEC in a tube shape potential was introduced in Refs.\,\cite{pal2020enf,Blakie2020vtf}. The basis of this theory is to decompose the 3D field as $\psi(\mathbf{x})=\sqrt{n}\chi_\mathrm{var}(\brho)$ where $\chi_\mathrm{var}(\brho)=\tfrac{1}{\sqrt{\pi}l} {e^{-(\eta x^2+z^2/\eta)/2l^2}}$ is a two-dimensional Gaussian function with variational parameters $\{l,\eta\}$, and  $n$ is the (uniform) linear density along $y$. The energy per particle is given by
\begin{align}
\mathcal{E}_u(l,\eta)  =\mathcal{E}_{\mathrm{tr}}+ \tfrac{1}{2}n \tilde{U}_\alpha(0)+\tfrac{2}{5}g_{\mathrm{QF}}n^{3/2},\label{Eq:Eu}
\end{align}
where
\begin{align}
\mathcal{E}_{\mathrm{tr}} = \frac{\hbar^2}{4ml^2}\left(\eta+\frac{1}{\eta}\right)+\frac{ml^2}{4}\left(\frac{\omega_x^2}{\eta}+\omega_z^2\eta\right),
\end{align}
is the single-particle energy associated with the transverse degrees of freedom, 
 \begin{align}
 \tilde{U}_\alpha(0)=\frac {2\hbar^2}{ml^2}\left(a_s+a_{dd}b_\alpha\right),\label{U0}
 \end{align}
 is the effective interaction, obtained by integrating out $\chi_\mathrm{var}$, and $g_{\mathrm{QF}}=\frac{256}{15\pi l^3}\frac{\hbar^2a_s^{5/2}}{m} \mathcal{Q}_5(\epsilon_{dd})$, with $\mathcal{Q}_5(x)=\Re\{\int_0^1\,du[1+x(3u^2 - 1)]^{5/2}\}$ \cite{Lima2011qfi}.  Here $\alpha=\{\parallel,\perp\}$ denotes the cases of dipoles parallel ($\hat{\mathbf{e}}=\hat{\mathbf{y}})$ or perpendicular ($\hat{\mathbf{e}}=\hat{\mathbf{z}})$ to the $y$-axis and $b_\parallel=-1$ and $b_\perp=(2-\eta)/(1+\eta)$. The parameters $\{l,\eta\}$ are determined by minimising Eq. \eqref{Eq:Eu}. \\ \indent
 For the excitations we need the $k_y$-dependence of the interaction which is given by
 \begin{align}
\tilde{U}_\alpha(k_y)= \tilde{U}_\alpha(0)
+ \frac{2\hbar^2a_{dd}}{ml^2}A_\alpha Q_{\alpha}^2e^{Q_{\alpha}^2}\text{Ei}(-Q_{\alpha}^2),\label{Uk}
\end{align} 
where $Q_\alpha^2=\frac{1}{2}q_\alpha(\eta)k_y^2l^2$,  
with $q_{\parallel}(\eta)=(\frac{2\eta}{1+\eta^2})^{4/5}$, ${A_\parallel=-3}$, and $q_{\perp}(\eta)=\sqrt{\eta}$, $A_\perp=3/(1+\eta)$.
The lowest excitation band can be computed employing the same-shape approximation, i.e.\,taking the excitations to have the same transverse profile as the condensate. From this treatment \cite{Blakie2020vtf} we obtain Eq.\,(\ref{eq:var_E}).

\section{Extended meanfield theory}\label{app:Theory}

We employ the extended Gross-Pitaevskii equation (eGPE) to model the ground states. Here the two-body interactions are described by the potential
 
\begin{align}
U({\bf r}) = \frac{4\pi a_s \hbar^2}{m} \delta({\bf r}) + \frac{3\hbar^2 a_{dd}}{m r^3} [1 - 3(\hat{\mathbf{r}}\cdot\hat{\mathbf{e}})^2].\label{Urfull}
\end{align}
The first term describes the short-ranged contact interactions with  $a_s$  being the s-wave scattering length. The second term describes the DDIs characterized by the dipole length ${a_{dd}=\mu_0\mu_m^2m/12\pi\hbar^2}$, with $\mu_m$  being the magnetic moment and $\hat{\mathbf{e}}$  being the dipole orientation direction. The eGPE equation for the ground state is
\begin{align}
\mu\psi_0=&\left(-\frac{\hbar^2\nabla^2}{2m}+V(\mathbf{x})+\Phi(\mathbf{x}) +\gamma_{\mathrm{QF}}|\psi_0|^3\right)\psi_0,
\end{align}
where $\mu$ denotes the chemical potential, ${V(\mathbf{x})=\frac{1}{2}(\omega_x^2x^2+\omega_y^2y^2+\omega_z^2z^2)}$ is harmonic confining potential, and $\Phi(\mathbf{x})= \int d\mathbf{x}'\,U(\mathbf{x}-\mathbf{x}')|\psi_0(\mathbf{x}')|^2$ is the interaction potential. The quantum fluctuations are described by the last term \cite{FerrierBarbut2016ooq,Waechtler2016qfi,Bisset2016gsp} with coefficient \begin{align}
\gamma_{\mathrm{QF}}=\frac{128\sqrt\pi\hbar^2}{3m}\sqrt{a_s^5}\mathcal{Q}_5(\epsilon_{dd})
\end{align}
For the infinite tube results (with  $\omega_y=0$) we solve for a ground state of  fixed linear density $n$ of the form $\psi_0(\mathbf{x})=\sqrt{n}\chi(\bm{\rho})$, with normalization condition $\int d\bm\rho\,|\chi|^2=1$ . Our results for the 3D harmonic trap are for a cylindrical symmetric potential ($\omega_\rho=\omega_{x,z}$) where the wavefunction reduces to the form $\psi(\rho,y)$, with normalization condition $N=\int\,2\pi d\rho\,dy\, |\psi|^2$. The ground states are found using an imaginary time algorithm. The quasiparticle excitations are found by solving the Bogoliubov-de Gennes equations, which can be obtained by linearizing the time-dependent eGPE around a ground state. The form of these equations for the dipolar eGPE are given in Ref.\,\cite{Baillie2017ceo} and are specialized to the infinite tube system in Refs.\,\cite{pal2020enf,pal2022idd}.

\section{Experimental Sequence}\label{app:ExpSequence}

\begin{figure}[ht]
	\centering
    \includegraphics[width=8.6cm]{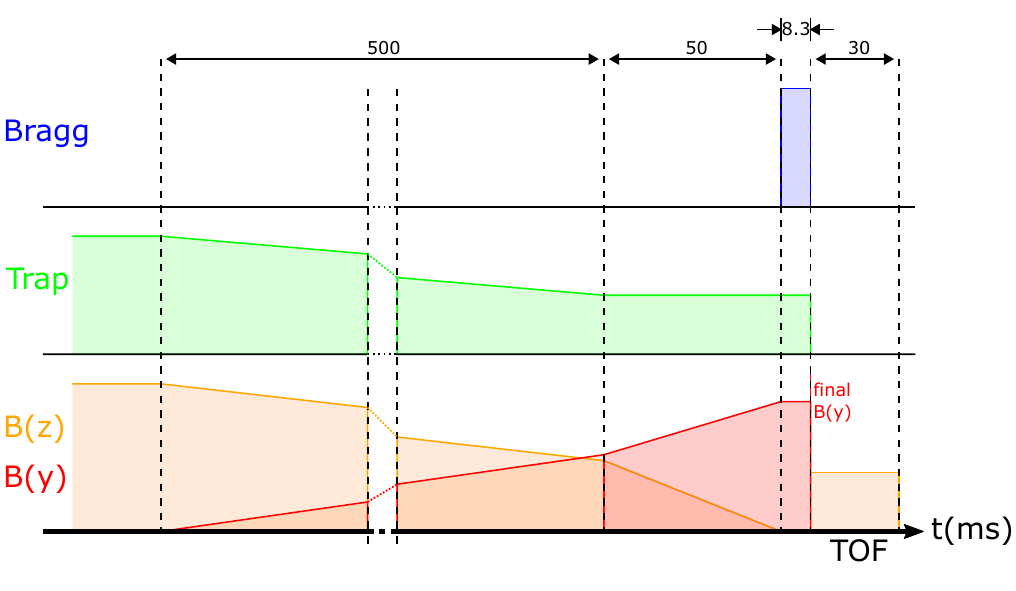}
     \caption{Sketch of the experimental sequence
     \label{appfig:ExpSequence}
    }
\end{figure}

A sketch of the experimental sequence can be found in Fig.\,\ref{appfig:ExpSequence}. To rotate the polarization into the long axis of the trap, the magnetic field in the $z$-direction is ramped down to $0.65\,$G while the magnetic field in the $y$-direction is ramped to the required value in $500\,$ms. In 50 additional milliseconds, the magnetic field in the $z$-direction is ramped to zero. Preparation in this way prevents high losses in the vicinity of the Feshbach resonance at $0\,$G and reduces breathing and center-of-mass modes to a minimum. 

To probe the prepared state, we perform Bragg-spectroscopy using a digital micromirror device (DMD, \textit{DLP9500 1080p VIS}).
Two gratings are uploaded onto the DMD and illuminated by one off-resonant beam. The two diffracted beams are then focused by a lens and interfere to form a lattice at the position of the cloud. By changing the phase of the gratings, the lattice can be made to move at a programmable velocity. The angle and hence the wavelength of the lattice can also be changed programmatically by changing the relative distance between the two gratings on the DMD. The phase step $d\phi$ is chosen to be small enough such that the measurement does not suffer from any higher harmonics in the frequency, yet large enough to be able to access the high frequencies needed to resolve multiple excitation branches. Since the DMD has a maximum refresh rate of $10.8$\:kHz, choosing $d\phi=2\pi/10$ limits the maximum excitation frequency therefore to $2\times10.8/10=2.16\:\mathrm{kHz}$ (the factor 2 coming from the frequency doubling of changing the phase of both gratings simultaneously). For a more detailed description of the setup, see the supplementary materials of Ref.\,\cite{Petter2019ptr}.

\section{Image Analysis} \label{appendix:imgAnalysis}

For a substantial number of our images we cannot reliably fit a Gaussian to the excitations because the excitations at low $k_y$ are either too close to or even partially within the unperturbed cloud, or the collisions between the scattered and unscattered particles distort and blur the excitation, see also \cite{veeravalli2008bss}. As described in the main text, we therefore choose to calculate $k_y$ numerically, relying on momentum conservation and not on any fitting parameters.

\begin{figure}[t]
	\centering
    \includegraphics[width=8.6cm]{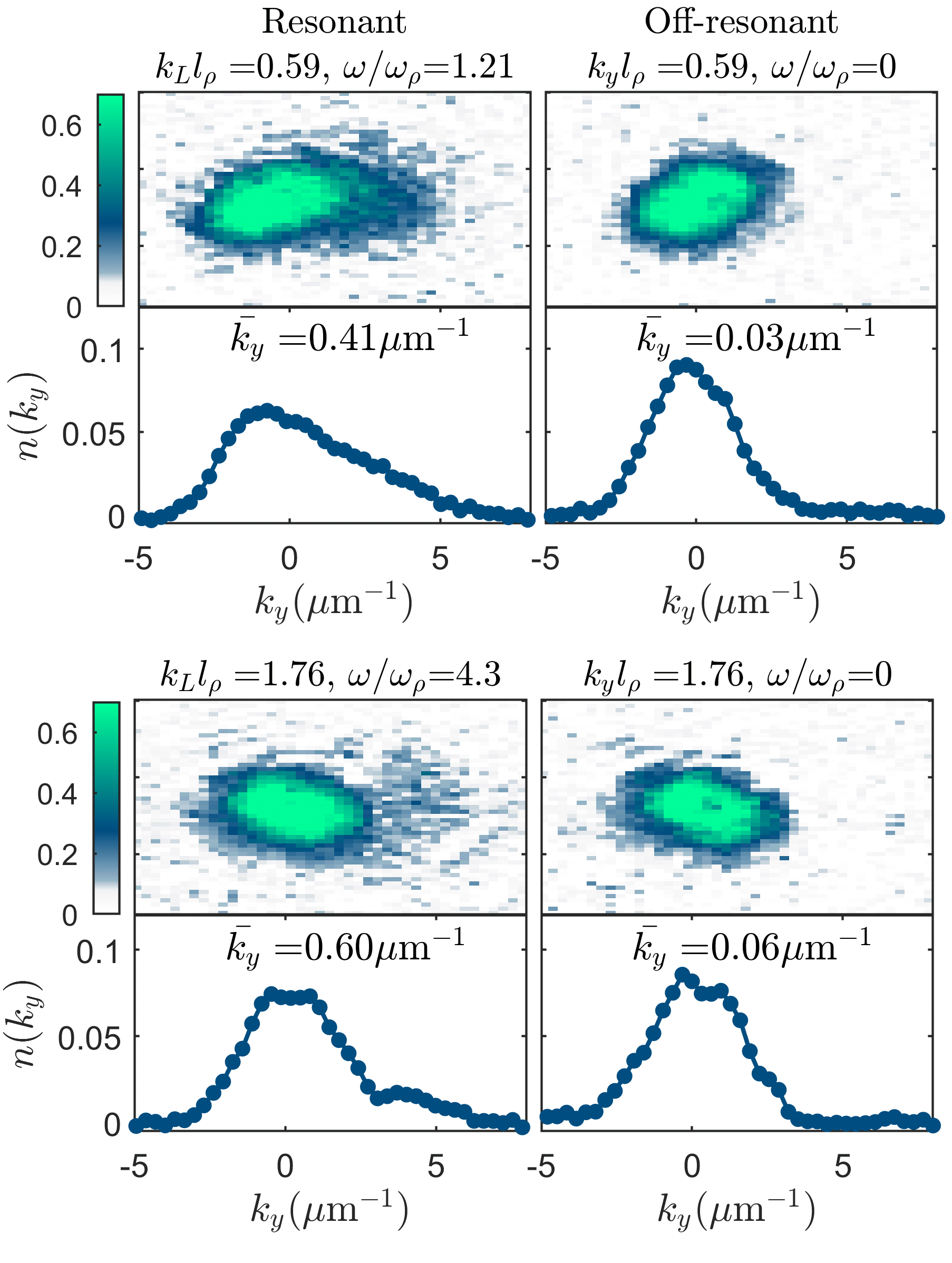}
     \caption{Examples of averaged absorption images taken on (left) and off (right) resonance after being cropped to the region of interest. The momentum distribution $n(k_y)$, normalized such that $\int n(k_y)dk_y\,=\,1$, and the corresponding calculated value of $\ky$ are shown in the lower plots. The top (bottom) images are examples of the measurements used in Fig.2b (Fig.2d).
     \label{appfig:imageAnalysisExample}
    }
\end{figure} 

To create a response for a fixed $k_L$, such as in Fig.\,\hyperref[fig:fig2]{\ref*{fig:fig2}b-d}, we take about 3-7 images per probe frequency $\omega_L$. Every image taken along the $z$-axis gives us, assuming ballistic expansion during TOF, the integrated momentum distribution of the cloud in the $x$-$y$ direction, see Fig.\,\ref{appfig:imageAnalysisExample} for two examples. For a TOF of $30\,$ms using our vertical imaging setup, the system has been calibrated with an optical lattice to resolve a momentum of $0.32\,\imicron$ per pixel. After subtracting the background and centering the image on the unperturbed cloud using a Gaussian fit, we crop the image to an asymmetric region of interest $k_x = \pm8\imicron$ and $k_y = [-5,5+2k_L]\imicron$, reflecting the asymmetry of the cloud and the fact that the excited atoms typically acquire $k_y \approx k_L$. We then average all images per probe frequency and integrate out $k_x$ to obtain the 1D momentum distribution $n(k_y)$. Finally, we calculate the mean momentum per particle by numerical integration $\ky=\int k_yn(k_y)dk_y/\int n(k_y)dk_y$, see Fig.\,\ref{appfig:imageAnalysisExample}.
To obtain the resonance positions of Fig.\,\hyperref[fig:expSpectrum]{\ref*{fig:expSpectrum}a-b}, a triple Gaussian is fit to all individual $\ky$ points in a measurement series.
Note that we sometimes observe a mismatch between the imprinted momentum $k_L$ and the peak in the momentum distribution of the diffracted atoms. We speculate that scattering events during the Bragg excitation pulse, interaction effects during the time-of-flight expansion, and residual higher-order or multiple Bragg excitations during the pulse time are modifying the observed momentum distribution.

\section{\texorpdfstring{Comparison between $S(\mathbf{k},\omega)$ and $\ky$}{Comparison between S(k,w) and ky}}\label{app:CompSandk}

\newcommand{\densfluctop}{\delta \hat{n}_{\mathbf{k}}}
For the specific case that the duration of the pulse $\tau$ is short with respect to the trap frequency $\omega_y=2\pi\nu_y$ (i.e. $\tau\nu_y \ll 1$), but long with respect to the excitation frequency $\omega$ ($\tau\omega/2\pi \gg 1$), it was shown in Ref.\,\cite{brunello2001mtt} [Eq.\,(14)] that the rate of change of momentum during a Bragg pulse can be related to the dynamic structure factor as 
\begin{equation}
   \frac{d\mathbf{k}(t)}{dt} = \mathbf{k}_L \frac{\pi V_0^2}{2 \hbar} S_0(\mathbf{k},\omega), 
\end{equation}
where $S_0(\mathbf{k},\omega) =S(\mathbf{k},\omega) - S(-\mathbf{k},-\omega)$ is the zero-temperature DSF~\cite{blakie2002tcb}, and where it is assumed that the initial position and momentum expectation values are zero.
In our case, with $\tau^{-1} = 120\,\mathrm{Hz}$ and $\nu_y = 30\,\mathrm{Hz}$, we roughly satisfy the above requirements when the excitation frequencies $\omega_L/2\pi > 120\,\mathrm{Hz}$ or $\omega_L/\omega_{\rho} > 0.6$, which is true for almost all measurements.
We can then directly relate our observable to the DSF as
\begin{equation}
   \ky =  k_L \frac{\pi \tau V_0^2}{2 \hbar}  S_0(k_L,\omega).
\end{equation}
A complication coming from the presence of multiple branches with comparable weight is that they all contribute to the imparted momentum. This effect is easier to appreciate if one recasts the integral in Eq.\,\eqref{eq:DSF} in terms of the density fluctuation operator
\begin{equation}
    \densfluctop = \int d\mathbf{x} \left(  \hat{\psi}^{\dagger}(\mathbf{x}) \hat{\psi}(\mathbf{x}) - \braket{0|\hat{\psi}^{\dagger}(\mathbf{x}) \hat{\psi}(\mathbf{x})|0}\right) e^{-i\mathbf{k}\cdot\mathbf{x}},
\end{equation}
which gives \cite{Zambelli2000dsf}
\begin{equation}
    S_0(\mathbf{k},\omega) = \sum_j |\braket{j|\densfluctop^{\dagger}|0}|^2 \delta(\hbar \omega - \epsilon_j).
\end{equation}
In practice, as the experiment does not probe the system with a delta function in $\omega$, one should rather consider the Fourier-broadened structure factor 
\begin{align}
   \nonumber \tilde{S_0}(\mathbf{k},\omega) &= \frac{1}{\hbar}[S_0(\mathbf{k},\omega')*\tau \sinc^2(\tau\omega'/2)](\omega)\\
    &= \sum_j |\braket{j|\densfluctop^{\dagger}|0}|^2 \tau \sinc^2(\tau(\omega - \epsilon_j/\hbar)/2),
\end{align}
which more accurately represents the frequencies probed by a moving lattice pulsed for a time $\tau$. To be even more precise which energies are probed, one should also include the Doppler broadening as described in App.\,\ref{app:Broadening}. If the separation in energy between two excitations is smaller than the total broadening, both excitations will contribute to the structure factor.

\section{Broadening mechanisms}\label{app:Broadening}

We identified Fourier-broadening and Doppler-broadening as our main experimental effects. In the following, the widths of the broadenings are calculated as $1 \sigma$ of a Gaussian, since in Fig.\,\hyperref[fig:expSpectrum]{\ref*{fig:expSpectrum}b} we use a Gaussian to broaden the numerical spectrum. For the Fourier-broadening we consider our pulse length of $\tau=8.3\,$ms, which corresponds to an energy broadening of $\delta\omega_F/2\pi\approx  47\,\Hz$, setting the lower limit to our resolution in energy. 

For the Doppler broadening we consider the Doppler shift of the excitation frequency seen from an atom moving with velocity $v_A$:
\begin{align*}
    \omega = (1+\frac{v_{A}}{v_L})\omega_L \\
\delta \omega = \omega - \omega_L = v_{A}k_L,
\end{align*}
where $\omega_L$ and $k_L$ denote the angular frequency and wavenumber of the lattice respectively, which in turn define the phase velocity $v_L=\omega_L/k_L$. A first contribution comes from the finite momentum width of the initial ground state. We estimate the expected Doppler broadening from the Fourier transform of the simulated ground state wavefunction, resulting in $\delta\omega_{D_D}/2\pi = 37\,\Hz \cdot k_L l_{\rho}$.
This calculation provides a lower bound to the Doppler broadening, since the thermal cloud surrounding the BEC in the experiment will most likely have an additional broadening effect. We also take into account the expected broadening from Doppler shifts arising from shot-to-shot fluctuations of the average momentum during the Bragg pulse. For this we measured the fluctuations in the central position of the cloud after TOF \cite{StamperKurn1999eop,Stenger1999bso,Steinhauer2002eso}, resulting in a Doppler-broadening of $\delta\omega_{D_S}/2\pi = 45\,\Hz \cdot k_L l_{\rho}$.

The total broadening is then calculated as the quadrature of the individual broadening mechanisms as
\begin{equation}
    \delta\omega(k_Ll_\rho)=\sqrt{\delta\omega_{D_D}^2+\delta\omega_{D_S}^2+\delta\omega_F^2},
\end{equation}
which -- for illustrative values of $k_yl_\rho=[1,2]$ -- evaluates to $\delta\omega(1)/2\pi=75\Hz$ and $\delta\omega(2)/2\pi=126\Hz$.
\vspace{5cm}